\documentstyle[psfig,preprint,aps,tighten]{revtex}
\preprint{\vbox{\baselineskip=12pt
\rightline{CGPG-97/1-2}
\rightline{gr-qc/9605047}}}

\def\B{{\cal B}}
\def\b{\beta}
\def\a{\alpha}

\begin{document}
\draft
\title{On Quantum Statistical Mechanics of \\
a Schwarzschild Black Hole}
\author{Kirill V. Krasnov\thanks{E-mail address: krasnov@phys.psu.edu}}
\address{Center for Gravitational Physics and Geometry, \\
The Pennsylvania State University, PA 16802, USA.} 
\date{\today}

\maketitle

\begin{abstract}
Quantum theory of geometry, developed recently in the framework
of non-perturbative quantum gravity, is used in an attempt to explain 
thermodynamics of Schwarzschild black holes on the basis of a microscopical 
(quantum) description of the system. We work with the formulation of 
thermodynamics in which the black hole is enclosed by a spherical surface 
$\B$ and a macroscopic state of the system is specified by two parameters: 
the area of the boundary surface and a quasilocal energy contained within. To 
derive thermodynamical properties of the system from its microscopics we 
use the standard statistical mechanical method of Gibbs. Under a certain
number of assumptions on the quantum behavior of the system,
we find that its microscopic (quantum) states are described by states 
of quantum Chern-Simons theory defined by sets of points on $\B$ labelled
with spins. The level of the Chern-Simons theory turns out to be 
proportional to the horizon area of black hole measured in Planck units. 
The statistical mechanical analysis turns out to be
especially simple in the case when the entire interior of $\B$
is occupied by a black hole. We find in this case that the entropy contained 
within $\B$, that is, the black hole entropy, is proportional to the horizon 
surface area. 
\end{abstract}
\pacs{Key words: black hole thermodynamics, quantum gravity}
\eject

\section{Introduction}
\label{sec:1}

The statistical mechanical method of Gibbs, developed in his
celebrated {\it Elementary Principles of Statistical Mechanics}
(1902), turned out to be the most efficient tool for 
explaining the bulk properties of matter on the
basis of its microscopics.
So far, self-gravitating systems have resisted the application of this
method. In this paper 
we apply the statistical mechanical method of Gibbs to such
systems. The aim of the paper is to try to derive thermodynamics of
gravitational systems using methods developed by Gibbs.

Our discussion is restricted to a simple gravitational system whose 
thermodynamics is widely studied: a spherically symmetric uncharged black hole.        
Although this paper deals only with a special example of gravitational
system, there is a natural way to generalize the ideas 
presented to other types of black holes (see \cite{ABCK}). 

As a basis of our discussion we take the thermodynamics of self-gravitating 
systems in the form it is formulated by Martinez \cite{Erik1}.
In this formulation the gravitational system composed of 
a spherically symmetric black hole 
is characterized by the surface area $A=4\pi R^2$ of a two-dimensional 
spherical boundary surface $\B$ (located at $r=R$) that encloses the black 
hole, and a quasilocal energy $E$ contained within. Following Martinez
\cite{Erik1}, we use the quasilocal energy of Brown and York \cite{BY}
as $E$. In equilibrium states 
the system is completely characterized macroscopically when 
these two variables are specified (see \cite{Erik1}). 

The expression of the entropy function $S$ as a function $S(A,E)$ 
of extensive variables is called fundamental equation of a 
thermodynamical system. 
Once known, the fundamental equation contains all
thermodynamical information about the system. For the system of our interest 
the fundamental equation is given by \cite{Erik1}
\begin{equation}
S(A,E) = 4\pi E^2 \left ( 1 - {E\over 2R} \right )^2,
\label{fundeq}
\end{equation}
where `radius' $R$ is used as an extensive variable instead of the area $A$.
Due to the spherical symmetry of the system 
one can use $R$ and $A$ interchangeably. 

The fundamental equation (\ref{fundeq}) is derived from 
Hawking's semiclassical expression \cite{Haw}
for the temperature of black hole radiation.
It is our aim in this paper to try to explain the fundamental equation
(\ref{fundeq}) on the basis of the microscopics of the system. To derive the
fundamental equation (\ref{fundeq}) we apply the Gibbs' method. 
The method of Gibbs, when adapted to our case, is to construct
a function $Q(\alpha, \beta)$ called the statistical sum of the system,
which is a function of intensive parameters of the system
(here $\alpha$ stands for the product $\beta p$ of the inverse temperature
$\beta$ and the quantity $p$ that plays a role of the `surface pressure'; see 
\cite{Erik1}). The statistical sum $Q$ is given by
\begin{equation}
Q(\alpha,\beta) = {\rm Tr}\, e^{-\alpha\hat{A}-\beta\hat{E}},
\label{statsum}
\end{equation}
where $\hat{A}$ and $\hat{E}$ are quantum mechanical operators 
corresponding to the classical quantities $A,E$. 
The statistical sum (\ref{statsum}) contains all thermodynamical
information about the system. In particular, the fundamental equation 
of the system can be obtained from $Q(\alpha,\beta)$ by means of
well-known thermodynamical relations. 

As a basis of microscopical description of our system we use the one given by 
non-perturbative quantum gravity based on the loop representation \cite{RS1}
(see also \cite{A1}). An important modification
arises, however, because our system has a boundary. 
In this paper we propose a 
construction of the space of quantum states which takes
into account the presence of the boundary.
Having found quantum states of the system, 
to calculate the statistical sum $Q$ one needs to 
construct quantum operators corresponding to the variables $A, E$, and
find the corresponding spectra. The latter problem in the case of the 
area operator has been resolved successfully in \cite{A1}. However,
there exists only rather tentative results concerning the quasilocal energy 
operator. 
Surprisingly, even without knowledge of the operator $\hat{E}$
it turns out to be possible to analyze the thermodynamics
of the system in a special case. Namely, we are able to complete the
statistical mechanical analysis in the case when the boundary 
surface $\B$ that encloses the system coincides with the horizon surface of
black hole. 
The analysis of this case constitutes the main result of the paper.

We wish to note that our work uses some of the ideas presented in 
earlier works \cite{GEntropy,Rov1,Rov2} on black hole thermodynamics
performed within the loop approach to quantum gravity. The important difference
between the present and earlier works is that we propose to use
Chern-Simons theory to describe states of the black hole quantum
mechanically. 

The organization of this paper is 
as follows. Section \ref{sec:2} describes the space of
quantum states of our system. Section \ref{sec:3} constitutes the core of the
paper. It contains a statistical mechanical analysis of the case 
when the whole interior of $\B$ is occupied by a black hole. 
We conclude with the discussion of the results obtained.
 
\section{Quantum States}
\label{sec:2}

In this section we describe physical quantum states of our system. 
We give only main points of the construction. 
For details see \cite{ABCK}.

Let us recall that our system consists of black hole of a certain mass 
$M$ enclosed by a spherical boundary surface $\B$ 
(located at $r=R, R \geq 2M$).
Macroscopically the system is completely characterized by two variables,
that is, the area $A$ of $\B$ and the quasilocal energy 
of Brown and York $E$ contained within,
that are both defined on $\B$. The quantities $A,E$ completely define 
(via vacuum Einstein's equations) 
the state of the gravitational field within $\B$.
Therefore, one can think of a macroscopic (classical) state of our system as of
a state of the gravitational field inside the surface $\B$.   

We shall describe microscopic (quantum) states  of the system as states
of the quantized gravitational field within $\B$. To construct these 
states we need to solve two problems, in complete analogy with the 
classical case. The first problem is to find the space of
kinematical states; these are analogs of `geometries'
of the classical case. The second problem consists in imposing `the
quantum Einstein's equations'. Solving these equations one finds the 
so-called  physical states.

A major progress  has been made on the first of these problems.
Following the pioneering work \cite{RS1} 
of Rovelli and Smolin, a mathematically
well-defined theory of `quantum geometry' has been constructed
(see \cite{A1,Rov2} for the most 
recent account). However,
a general treatment of `the quantum Einstein equations'
is still a matter of much controversy, despite the recent progress
on this front \cite{T}.

In this paper we propose a solution to the second problem for 
our special case, and find the physical states.
To describe the main idea of our treatment we need a simple observation 
from the classical theory.
Let us note that in the classical case the geometry within $\B$
is completely determined (via vacuum Einstein's equations) once the geometry
on the surface $\B$ is specified. Indeed, for our 
special case of static Schwarzschild spacetime
dynamics (in the sense of the Hamiltonian formulation) is trivial.
Therefore, the problem of solution of Einstein's equations reduces to
the problem of solution of constraints, which is an elliptic problem.
This means that, to find the geometry within $\B$, one should specify
an (allowable) geometry on $\B$ (or, so to say, boundary conditions). 
This determines (via constraint equations) the geometry within the entire $\B$.

Following the spirit of the correspondence principle advocated by N.Bohr 
one can expect that, whatever precise form of equations of quantum 
dynamics is, these equations will keep some key properties of their
classical analogs. For example, it is natural to expect that  
solutions to quantum dynamical equations describing static 
(in some appropriate sense) geometry within $\B$ should be 
completely determined by a state of geometry on the boundary $\B$,
as this happens in the classical case. We  assume that 
(still unknown) equations of quantum dynamics possess this property.
The assumption implies that physical states of our
system are labeled by quantum states of geometry on the boundary
surface $\B$.
Note that our assumption does not tell us anything about 
non-stationary cases for which the corresponding classical property
does not hold.

Thus, the analogy with the classical case tells us that equations
of quantum dynamics should be such that, once a
state of quantized gravitational field on $\B$ is specified,
one finds a state of quantized gravitational field within the
entire $\B$ solving these equations. However, even in the 
classical case there does not exist a solution for any choice of
`boundary data'. We expect, therefore, that in the quantum case
not all states of quantum geometry on the surface $\B$ will give
rise to physical states of the system. Moreover, in the classical case
we need to know only certain minimal set of data to find the geometry
within $\B$. We expect to have an analog of this in the quantum case.

Thus, to construct quantum states of the Schwarzschild black hole
let us discuss the classical case in more details, and find a minimal
set of surface data that completely specifies a macrostate of the system. 
We describe the gravitational field using 
Ashtekar variables \cite{A2}. The geometry of spatial hypersurfaces
$\Sigma_t$ is described in this formalism by a pair of (canonically
conjugated) variables $\tilde{E}^a, A_a$. Let us now introduce
fields describing the geometry on $\B$. First, we introduce the
pullback $a$ of the connection $A$ on $\B$. Next, let us denote
by $e$ the two-form that is the pullback on $\B$ of the dual
to the densitiezed triad $\tilde{E}^a$ two-form
$\epsilon_{abc}\,\tilde{E}^c$.

On solutions of Einstein's equations the surface fields $a,e$ are
not independent. In the Schwarzschild spacetime there exists
a simple relation between the curvature
two-form of the connection $a$ and the two-form $e$. Taking from \cite{Ben}
explicit expressions for $a,e$ in the Schwarzschild spacetime,
one can find that
\begin{equation}
f = {2M\over R}\,{1\over R^2}\,e.
\label{relation}
\end{equation}
Note that throughout this paper we use the units in which $\hbar=G=c=1$.

We take (\ref{relation}) as the sought
surface constraint equation. Thus, we view two fields
$e, a$ on the surface $\B$, where the connection $a$ is such 
that its curvature two-form is related to $e$ via $f = \psi e$, 
$\psi$ being a number, as specifying a
macroscopic state of our system. Indeed, it is easy to see that
classically this set of data carries all information necessary to reconstruct
the geometry within the entire $\B$. Let us note that the
two form $e$ carries information about areas of regions on $\B$,
and, in particular, the area of the entire $\B$ can be determined
once $e$ is known. If, moreover, one is given that the curvature
two-form $f$ of the connection on $\B$ is related to $e$ via
$f = \psi e$, one can find the mass of the corresponding
Schwarzschild solution from (\ref{relation}). 
Thus, a set of fields $e, a$ together with the relation
(\ref{relation}) between them carries all information about a
macrostate.

Following our method of analogy with the classical case, we now
simply have to find how to describe states of 
fields $e, a$ quantum mechanically, and how to impose the constraint
(\ref{relation}) in the quantum case. The states we find this way 
will label the physical states of our system.

We are lucky, for there exists a complete description of states we look
for. Quantum theory of geometry \cite{A1} provides us with a 
complete description of states of quantized $e$ field. Conformal
field theory \cite{Conf} tells us what are quantum states of 
connection subject to a quantum analog of constraint (\ref{relation})
on a two-dimensional surface. The corresponding quantum states,
although in a different context, have already been used in
non-perturbative quantum gravity by Smolin \cite{Linking}. In this paper
we simply give a description of these quantum states in the amount
we need for our statistical mechanical treatment. 

\begin{figure}
\centerline{\hbox{\psfig{figure=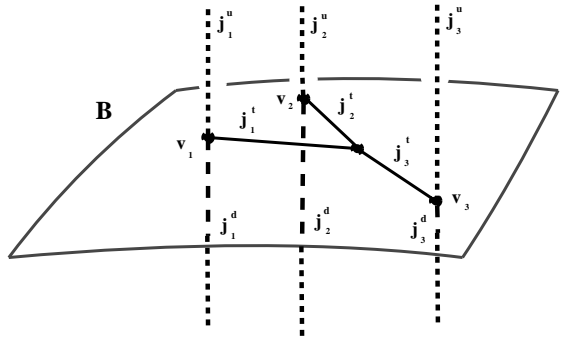}}}
\caption{A typical quantum state of our system
is labeled by a set of vertices, with 
quantum numbers (spins) attached. The spins $j$ should satisfy 
the condition $j \leq k$. 
For every set of vertices on $\B$ (and quantum numbers)
there exists only a finite number of states of quantum connection,
which can be thought of as the number of different ways in which 
the spins $j_v^t$ can be combined consistently according to
the rules of addition of angular momentum of the quantum group $SU(2)_q$.}
\label{fig:1}
\end{figure}

The states we are going to describe are invariant under diffeomorphism
transformations that are tangent to the boundary surface $\B$, 
gauge invariant, and satisfy, in some precise sense, the quantum 
analog of the constraint (\ref{relation}). 
We give a description of a basis in the
space of quantum states that is formed by eigenstates of operators measuring
areas of regions on $\B$. This basis is especially convenient for
our purposes.
A basis quantum state is labeled by a
set of points on the surface $\B$, which we, following \cite{A1},
shall call {\it vertices}. It is important that 
two states that differ only in position of
vertices on $\B$ should be considered as a single state if one
deals with diffeomorphism invariant states. Each vertex $v$ is labeled
by a set of quantum numbers. These include: (i) spin (half-integer) $j_v^{d}$
that enters the interior of $\B$ starting at the vertex $v$;
(ii) spin $j_v^{u}$ that enters the surface
$\B$ on an edge lying up the surface;
(iii) spin $j_v^{t}$ that stays on the surface (see Fig. \ref{fig:1}). 
Because of gauge invariance
quantum number $j_v^t$ can take values only from the range 
$|j_v^u - j_v^d|\leq j_v^t \leq j_v^u + j_v^d$. Note also, that
because of gauge invariance not all sets of spins are allowed
\cite{A1}. Namely, in the case of a closed surface, which we consider
here, the half-integers $j_v^u, j_v^d$ should satisfy the
additional condition that the sums $\sum_{v\in\B} j_v^u$ and 
$\sum_{v\in\B} j_v^d$ over all vertices in $\B$ are integers.

The set of quantum numbers we have described can be thought of as
specifying a state of $e$ field on $\B$. As it is described by
conformal field theory, for each choice of these
quantum numbers there exists a finite number of states of
the quantized connection $a$ on $\B$ that satisfy the constraint 
(\ref{relation}). These states are the states of quantum Chern-Simons
theories defined by a set of spins $\{ j_v^t \}$.
Different possible states of the quantized connection $a$
are labeled by additional quantum numbers.   

It turns out (see, for instance, \cite{Witten} for a discussion of this
point) that the zero vector is the only vector in the physical 
Hilbert space of states of quantized connection $a$, unless the representations
of $SU(2)$ labeled by spins $j_v^t$ all satisfy a certain condition of 
integrability. It is shown, for example, in \cite{Linking} that
for the case when the constraint equation is 
\begin{equation}
e = {k\over 4\pi} f,
\label{q1}
\end{equation}
the integrability condition is simply that all spins $j_v^t$ satisfy
the inequality $j_v^t\leq k$. Here $k$ is required to be an integer.
Comparing (\ref{q1}) with the relation (\ref{relation}) we find that
in our case
\begin{equation}
k = A\,{1\over (2M/R)}.
\label{level}
\end{equation}

It is known that 
representations satisfying the above integrability condition are,
in fact, representations of the quantum group $SU(2)_q$.
The number
$k$ is known as level, or coupling constant of the corresponding 
Chern-Simons theory \cite{Linking}. It is related in a simple way to the
deformation parameter $q$ (see, for example, \cite{Linking}).

It is not hard to see, that the level $k$ takes the minimal 
(for a fixed area $A$) value
$A$ when the surface $\B$ coincides with the horizon surface of
black hole. When the mass of black hole inside $\B$ decreases,
the level increases, as it is obvious from (\ref{level}). Thus, the
level $k$ is proportional to the area of the surface $\B$ (measured
in Planckian units) when this surface coincides with the horizon 
surface of black hole, and this value is the minimal possible value
once $A$ is fixed. This has an important consequence to us, for
this means that to large (in Planckian units) black holes correspond
large levels $k$. It is well-known that in the limit of large $k$
the dimensions of the quantum group representations go into dimensions
of their classical analogs. Also, there exists a simple formula
for the number of different states of quantized connection in the
limit of large $k$. Roughly speaking, a basis in the space of 
states of quantum connection $a$ is labeled by different ways
that the spins $j_v^t$ can be combined consistently according to
the rules of addition of angular momentum of the quantum group $SU(2)_q$.
In the limit of large $k$ this number is given by (see \cite{Linking})
\begin{equation}
\prod_v (2 j_v^t + 1).
\label{degeneracy}
\end{equation}

One can expect that statistical mechanical methods can only be
applied to macroscopical objects that are composed of many `elementary' 
excitations. For our geometrical system this means that statistical
mechanical description is presumably legitimate only when applied
to systems large as compared with Planckian scales. As we have seen,
for large black holes the level $k$ of the quantum theory is large,
and one can use the simple formulas that hold in the limit of a large $k$.
Thus, we shall not keep track of the fact 
that $k$ is finite. This can be shown to be legitimate, for the
corrections that appear when one takes the finiteness of $k$ are
negligible.

The states we have described (see Fig. \ref{fig:1} for a picture of a
typical quantum state) are eigenstates of operators that 
measure areas of regions on $\B$. In particular, the eigenvalues
of operator $\hat{A}_{\cal R}$ that measures the area of a region $\cal R$ 
on $\B$ are given by \cite{A1}
\begin{equation}
A_{\cal R}({\cal S}) = 16\pi\gamma \sum_{v\in{\cal R}} {1\over2} 
\sqrt{2j_v^u(j_v^u+1) + 2j_v^d(j_v^d+1) - j_v^t(j_v^t + 1)},
\label{qarea}
\end{equation}
where we have introduced the notation $\cal S$ for the basis quantum states.
The sum here is taken over all vertices of $\cal S$
that lie in the region $\cal R$.
We have taken into account in (\ref{qarea}) the fact that states
$\cal S$ are gauge invariant. The quantity $\gamma$ in (\ref{qarea}) is
the parameter that appears in the loop quantization of general relativity,
as it is discussed in \cite{Amb}.
In the case of closed surface, which we consider here, not all
of the eigenvalues (\ref{qarea}) are eigenvalues of the operator
$\hat{A}$ that measures the area of the entire $\B$. Namely, in the case
when $\B$ is closed gauge invariant quantum states are those that satisfy
the additional condition that we mentioned above.

Unfortunately, there does not exist yet such a complete description
of operator $\hat{E}$ that measures the quasilocal energy. However,
as we shall see in the next section, 
a partial statistical mechanical analysis is possible
even without knowledge of $\hat{E}$. 

\section{Statistical Mechanical Analysis}
\label{sec:3}

The statistical mechanical method of Gibbs consists in constructing
the statistical sum over all different quantum states of the system.
Since a macroscopic state is specified by the two extensive parameters:
the area $A$ and the quasilocal energy $E$, we have to introduce
two conjugate intensive parameters that we shall denote by $\alpha,
\b$ correspondingly. The statistical sum is a function of 
the intensive parameters
\begin{equation}
Q(\alpha, \b) = {\rm Tr}\, e^{-\a\hat{A}-\b\hat{E}}.
\label{s1}
\end{equation}

The statistical
sum can not be calculated, unless the operator
$\hat{E}$ is known. Let us note, however, that for two values of the
parameter $\b$, namely, for $\b = 0, \infty$, the calculation is
possible. Indeed, for $\b = 0$ a precise form of the operator $\hat{E}$
simply does not matter. In the latter case only the states of zero 
energy survive in (\ref{s1}).

It is obvious that the latter case describes the zero energy state 
of the system 
(it corresponds to the zero temperature, that, as we know, 
is proportional to the inverse $\b$).
In other words, in the case $\b=\infty$ our system is simply 
a region of flat spacetime enclosed by the surface $\B$. When one
decreases $\b$ (increases the temperature) the mean value of energy
$E$ will increase. One finds that the value $\b = 0$
(infinite temperature) corresponds to the largest possible value
of energy for fixed $A$. But, we know that the maximal (for a fixed area $A$)
possible `amount' of quasilocal energy is contained within $\B$
when all the interior of $\B$ is occupied by a black hole. Thus, the 
case of $\b = 0$ will describe the situation when $\B$ coincides
with the horizon surface of black hole. This is in agreement 
with the thermodynamics described
in \cite{Erik1}. Indeed, $\b$ becomes zero when the surface $\B$ 
coincides with the horizon surface of black hole (see \cite{Erik1}). 
Thus, we find that some
thermodynamical functions describing black hole can be calculates
even without knowledge of $\hat{E}$. To calculate thermodynamical
functions in this case, we simply have to put $\b = 0$ in (\ref{s1}).

To find $Q(\alpha,\b)$ in the case $\b = 0$ let us use the basis of
eigenstates of $\hat{A}$ described in the previous section. 
It is convenient to think about vertices on the surface $\B$ as about 
imaginary particles. Different
particles correspond to different values of quantum numbers
$j^u, j^d, j^t$. There can be any number of particles of each sort
in a state. It is important to take into account the fact that,
according to formula (\ref{degeneracy}), each particle `carries'
the degeneracy $(2j^t+1)$ (that is, the addition to a state of a
particle increases the dimension of space of states of quantized connection).
The sum over all states becomes the sum over numbers of particles 
\begin{equation}
Q(\a) = \prod_{\Gamma} \sum_{\{n_\Gamma\}} 
(d(\Gamma))^{n_\Gamma}\exp{\left( -\a\,n_\Gamma\,A(\Gamma)\right)} = 
\prod_{\Gamma} {1\over 1 - d(\Gamma)\exp{\left(-\a\,A(\Gamma)\right)}},
\label{s2}
\end{equation}
where we have denoted by $\Gamma$ a sort of particles ($\Gamma$ simply labels
a set of spins $j^u,j^d,j^t$), by $d(\Gamma) := (2j^t+1)$ the degeneracy
carried by a particle of sort $\Gamma$, and by $n_\Gamma$ the number
of particles of sort $\Gamma$ in a state. The summations over $n_\Gamma$
run from zero to infinity, $A(\Gamma)$ denotes the contribution 
to the area of $\B$ from a particle of sort $\Gamma$.

For simplicity, we have not taken into account in (\ref{s2}) the fact
that not all eigenvalues (\ref{qarea}) are eigenvalues of area operator
for a closed surface. In our language of particles this means that not
all combinations of particles can be realized. Thus, strictly speaking,
the summations over numbers $n_\Gamma$ of particles in (\ref{s2})
are not independent. Let us forget about this for a moment and 
proceed with our analysis. We shall discuss the consequences of this
restriction later on.

We have found that the state of the system 
when the entire interior of $\B$ is occupied by a black hole corresponds
to $\b = 0$. Thus, such a state of the system is described by a single
thermodynamical parameter $\a$. The mean values of the area $A$ and the
energy $E$, that can be expressed as derivatives of $\ln{Q(\a,\b)}$ at 
$\b=0$, become functions
of this single parameter. One could exclude $\a$, and express all
thermodynamical functions in terms of, for example, $A$. In other words,
a macroscopic  state of a Schwarzschild black hole is specified 
by a single parameter.
We are particularly interested to find the entropy function of the system
as a function of $A$. Clearly, it may serve as a first test to our
method whether it predicts the Bekenstein-Hawking formula for the
entropy of black hole. Namely, the thermodynamics described in 
\cite{Erik1} tells us that, when the whole of $\B$ is occupied by a black hole,
the entropy `contained' within $\B$ is simply the Bekenstein-Hawking 
entropy of black hole
\begin{equation}
S_{BH} = {1\over 4} A
\label{BH}
\end{equation}

To find the entropy as a function of $A$ one should, in principle,
find $S(\a)$ and $A(\a)$.
Then, one would exclude $\a$ and find $S$ as a function of $A$.
This problem can, in principle, be solved; however, there exists a 
simpler way to get the dependence $S(A)$. We note that
we expect the Bekenstein-Hawking formula (\ref{BH}) to hold
only for large (as compared with Planckian area) areas $A$.
Indeed, from the point of view of quantum geometry
only large as compared with the Planckian scale black holes
can be considered macroscopical, and thus, the thermodynamical description
can be applied only to such black holes. Thus, we are interested
in the dependence $S(A)$ predicted by the statistical mechanics only
for large values of $A$. 
One can change the mean value of area changing
the intensive parameter $\alpha$. Some values of $\a$ correspond
to large mean values of $A$, and we are to find  the
thermodynamical functions for these $\a$.

From the formula 
\begin{equation}
A(\a) = -\,{d\ln{Q(\a)}\over d\a}
\end{equation}
we find for the mean value
$A(\a)$ of the area 
\begin{equation}
A(\a) = \sum_\Gamma A(\Gamma) f(\Gamma),
\label{s3}
\end{equation}
where we have introduced the function
\begin{equation}
f(\Gamma) = {d(\Gamma)\exp{\left( -\a\,A(\Gamma)\right)} \over 
1 - d(\Gamma)\exp{\left( -\a\,A(\Gamma)\right)}}.
\label{s4}
\end{equation}
It is easy to see from (\ref{s3}) that $f(\Gamma)$ plays the role 
of the mean number of particles in the `state' $\Gamma$ for fixed value
of parameter $\a$. Large values of $A(\alpha)$ correspond to a case
when at least for some $\Gamma$ the function $f(\Gamma)$ takes large
values.

It can happen that the second term in the denominator of (\ref{s4}) 
for some $\Gamma$ is close to unity. This will correspond to a large
number of particles of the sort $\Gamma$, and, therefore, to a large
mean area.
Let us see to what values of $\a$ this corresponds. 

As it was noted in \cite{GEntropy}, there exists only few 
possibilities. First, it can happen that the thermodynamical
functions such as $Q(\a), A(\a), S(\a)$ get large values only when $\a\to 0$.  
It follows then from the Euler relation that in the thermodynamical
limit (large areas) the dependence $S(A)$ of the entropy on the area
is different from the linear one (in fact, $S(A)$ grows slower than
$A$). The second case is when the thermodynamical functions diverge
as $\a$ goes to some `critical' value $\a_{cr}$. Then, as it is discussed
in \cite{GEntropy}, the Euler relation implies the linear dependence
$S(A) = \a_{cr}\,A$. Finally, there exists the possibility that 
the statistical sum $Q$ diverges for all $\a$, which means that the
density of states of the system grows faster than $\exp{A}$.

It is not hard to show, that in our case the second possibility is
realized, and, therefore, the dependence $S(A)$ of the entropy on 
the area (for $\b=0$) is  linear. Indeed, let us consider the
function $d(\Gamma)\exp{-\,\a\,A(\Gamma)}$  as a function of $\Gamma$ and $\a$.
Let us recall that $\Gamma$ stands for three quantum numbers $j^u,j^d,j^t$.
It turns out to be convenient to use instead of $j^u,j^d,j^t$ the 
following linear combinations of them
\begin{eqnarray}
I := j^u+j^d, \nonumber \\
J := j^u-j^d, \label{newlabels} \\
K := j^u+j^d-j^{u+d}. \nonumber
\end{eqnarray}
Then, as it is not hard to see, $I,J$ take all positive integer
and half-integer values, and $K$ runs from zero to $2{\rm min}(j^u,j^d)$
($K$ also takes both integer and half-integer values). The contribution to
the area from a particle of sort $\Gamma$ written as a function of 
this new quantum numbers becomes
\begin{equation}
A(\Gamma) = A(I,J,K) = 8\pi\gamma\sqrt{J^2+I+2IK+K-K^2}
\end{equation}
The degeneracy that is `carried' by a particle of sort $\Gamma$ becomes
$2(I-K)+1$. As a straightforward analysis shows, for all
values of $\a < \a_{cr} = 1.138/8\pi\gamma$ the second term in the denominator
of (\ref{s4})
is less than unity for all types of particles $\Gamma$ (for all
non-zero values of quantum numbers $I,J,K$). When $\a\approx\a_{cr}$,
the maximal value of the function $d(\Gamma)\exp{-\,\a\,A(\Gamma)}$ 
in the allowable range of the quantum numbers $I,J,K$ is just
the unity. This is realized for the following values of quantum numbers:
$j^u=j^d; j^u+j^d=j^t=2$. 

Thus, what we find is that for the value of $\a$ close to $\a_{cr}$
the number of particles $f(\Gamma)$ in one of the `states' 
$\Gamma=\Gamma_{cr}$ becomes large. This means that the thermodynamical 
functions diverge for $\a=\a_{cr}$. As we have said above, this implies 
that for large values of $A$ the dependence of the
entropy on the area is linear
\begin{eqnarray}
S = \a_{cr}\,A, \nonumber \\
\a_{cr} = 1.138/8\pi\gamma 
\label{result1}
\end{eqnarray}
Note that we have obtained this result analyzing the behavior of
$f(\Gamma)$ as a function of $\Gamma$ and $\a$. It is crucial for the
result that our particles `carry' the degeneracy $d(\Gamma)$, for it
is this degeneracy in (\ref{s4}) that makes the number of particles 
in one of the states diverge for a finite (non-zero) value of $\a$. 

As we have said, when $\a$ approaches $\a_{cr}$ the  
number of particles in one of the states $\Gamma=\Gamma_{cr}$
goes to infinity.  It is interesting to compare this with the 
phenomenon of Bose-Einstein condensation. In the case of 
an ideal Bose gas a macroscopic fraction of 
particles that compose the gas end up on the
ground energy level when the temperature of the gas reaches certain
critical temperature. Let us note that
the number of particles in the case of a Bose gas is fixed. In our case
particles are imaginary creatures, and their number is not fixed.
As we have found, when the intensive parameter $\a$ approaches the
critical value $\a_{cr}$, much larger number of particles end up
in the state $\Gamma_{cr}$ than in all other states. 
Thus, what we find is very similar to Bose-Einstein condensation;
the important difference is, however, that the number of particles 
in our `gas' is not fixed. Indeed, the fact that the number of particles
is not fixed can be shown to be crucial for our result of linear 
dependence of the entropy on the area to hold.\footnote{%
A similar phenomenon, when the entropy of the system grows
linearly as a function of energy, is known to occur in ordinary
thermodynamics for systems for which the number of particles is not
fixed. The author is grateful to P. Aichelburg from whom he learned 
about this fact.}

When $\a$ is close to $\a_{cr}$ only a negligible fraction of particles
is in states different from $\Gamma_{cr}$. Thus, one can write
\begin{equation}
A = A(\Gamma_{cr})\,N,
\label{s5}
\end{equation}
where we have introduced the number $N$ of particles in the state
$\Gamma_{cr}$. 

Let us now recall that all results of this section were obtained 
neglecting the fact that, in the case of a closed surface $\B$,
not all sets of spins on $\B$ correspond to physical states. As we
have mentioned above, there exists a simple condition on the total
sums of spins that enter and leave the surface. The way to take into account
this condition is to exclude from the statistical sum all terms with 
spins not satisfying the conditions. Generally, such an operation
may significantly change the behavior of all thermodynamical functions that are
derived from the statistical sum. It is not hard to see, however, that
our condition is very weak in the sense that the results we have
obtained for the case of an open surface continue to hold when 
one considers closed surfaces $\B$. Here we present only a
heuristic argument in the support of this, 
leaving a more rigorous treatment for another occasion.

It is not hard to see that the
results (\ref{result1}), (\ref{s5}) stem from one
and the same fact that we have discovered about the microscopic state
describing black hole. We saw that a state describing a large
black hole is very special in the sense that a macroscopic fraction 
of `particles' resides in one and the same `state'. Then the total area
is given simply by the number of particles times the contribution 
from an individual particle, as it is described by (\ref{s5}). The entropy,
on the other hand, is given by the logarithm of the number of states
of quantized connection 
\begin{equation}
S = \ln{d(\Gamma_{cr})^N} = N\,\ln{d(\Gamma_{cr})}.
\end{equation}
Combining this equation with (\ref{s5}) we find that 
\begin{equation}
S = {\ln{d(\Gamma_{cr})}\over A(\Gamma_{cr})}\,A =
{\ln{5}\over 8\pi\gamma\,\sqrt{2}}\,A = \a_{cr}\,A.
\end{equation}
Since $j^u,j^d$ in this state
happen to have one and the same spin $1$,
the conditions we have mentioned above are satisfied in  any state
describing a large black hole. Thus, in the thermodynamical 
limit (large black holes)
the results we have obtained continue to hold even when one imposes the
conditions arising in the case of a closed  $\B$.

\section{Discussion}
\label{sec:6}

We have found
that the microscopic states of Schwarzschild black hole can be
described by states of ${\rm SU(2)}$ Chern-Simons theory,
which are defined by choices of vertices and spins on $\B$. We have found
that the integer level of Chern-Simons theory is proportional to the
horizon area of black hole (see (\ref{level})). Thus, large black holes
correspond to large levels $k$. We have used this description as the
basis of our statistical mechanical analysis. 

Although we have not reached our goal of explaining the fundamental
equation (\ref{fundeq}) on the basis of quantum microscopics of the 
system, we were able to complete the statistical mechanical
analysis of the case when the entire interior of the system is occupied by a
black hole. We have found that in this case the entropy contained
within $\B$ is proportional to the area of the boundary $\B$, with 
the proportionality coefficient given by (\ref{result1}). The 
proportionality coefficient between black hole entropy $S$ and the
horizon area $A$ turns out to be a function of the parameter $\gamma$. 

As it is discussed in \cite{Amb}, the parameter $\gamma$ is a free parameter
that arises in the loop quantization of general relativity. Unless
a value of the parameter $\gamma$ is fixed by some independent 
considerations, a comparison of the dependence $S(A)$ predicted by our
analysis with the Bekenstein-Hawking formula (\ref{BH}) is not possible.
Thus, the predicted dependence (\ref{result1}) by itself does not provide
us with a test of our approach. Note, however, that the approach presented
gives us not just the dependence (\ref{result1}). The considerations of
section \ref{sec:3} led us to the conclusion that the statistical state of
a black hole should be described by a density matrix
$$\hat{\rho} = {1\over Q(\a)} e^{-\a\hat{A}},$$
where $Q(\a)$ is the statistical sum (\ref{s1}). The entropy was then 
defined as $S = {\rm Tr}(\hat{\rho}\ln{\hat{\rho}})$. However, with 
the density matrix $\hat{\rho}$ in one's disposal one knows much more than 
simply the entropy. For example, the knowledge of the density matrix 
allows one to analyze properties of the black hole radiation spectrum  
\cite{Rad}. Results of such an analysis, 
together with the result (\ref{result1}), can serve
as a test of the validity of the approach presented.

The other result of this paper is the discovery of the fact
that the state of the system that corresponds to a macroscopically
large black hole is realized for values of the intensive parameter $\a$
that are close to the critical point $\a_{cr}$. This is quite similar
to what one finds, for example, in the sum over lattices approach to quantum
gravity in two dimensions \cite{2D}. One finds
that, in order to go into a macroscopical limit, one needs to tune
values of the parameters to a critical point. It is interesting
that in our case, when the parameter $\a$  goes to a critical point,
the system undergoes a phase transition similar to the phenomenon
of Bose-Einstein condensation. We find that a macroscopical fraction
of elementary excitations of geometry is in one and the same quantum state
when we are dealing with large black holes. The state where most of
the particles `condense' to happen to have the following quantum numbers:
$j^u=j^d=1$. 

Let us conclude summarizing the open problems of our approach. One of the most
important of such problems is to develop a theory of quantum dynamics of
geometry, and to answer the question whether the assumptions about dynamics
made in this paper are true. This is, however, a difficult problem that
may require mutual efforts of a large number of researchers in the field.

Probably a simpler, but not less important problem is to construct
the operator $\hat{E}$ corresponding to the quasilocal energy.
This would allow one to analyze the case of an arbitrary mass contained 
within $\B$ and find the fundamental equation of the system. 
Work is in progress in this direction.

Another important open problem, related to the problem of construction
of the energy operator, is to understand how the different mathematical 
techniques used in conformal field theory and in non-perturbative
quantum gravity can be reconciled in a mathematically rigorous construction
of quantum states. Work is in progress also in this direction
\cite{ABCK}.

\section{Acknowledgments.}

I am grateful to A.Ashtekar, A.Coricci, S.Major, C.Rovelli, L.Smolin,
J.Zapata and especially R.Borissov and E.Martinez  for discussions,
comments and criticism. 
This work was supported, in part by the International
Soros Science Education Program (ISSEP) through grant No. PSU062052,
by the NSF grant PHY95-14240 and by Eberly research Fund of Penn
State University. The author is also grateful for the support
received from the Erwin Schr\"odinger Institute for Mathematical
Sciences, Vienna.

\end{document}